\documentclass[aps,prl,reprint,groupedaddress]{revtex4-1}
\usepackage{graphicx} 
\usepackage{dcolumn} 
\usepackage{bm}
\usepackage{hyperref}

\begin{document}

\title{Nuclear in-medium effects and collective flows in heavy-ion collisions at intermediate energies}

\author{Zhao-Qing Feng}

\altaffiliation{Present address: Institut f\"{u}r Theoretische Physik, Justus-Liebig-Universit\"{a}t Giessen, Germany}
\affiliation{Institute of Modern Physics, Chinese Academy of Sciences, Lanzhou 730000, People's Republic of China}

\date{\today}

\begin{abstract}
A recent updated version of the Lanzhou quantum molecular dynamics is reviewed, in which the momentum dependence of symmetry potential and the effective mass splitting of proton and neutron in nuclear medium are included in the model. The in-medium nucleon-nucleon (NN) elastic cross sections are evaluated with the scaling approach according to the effective mass and their influence on collective flows in heavy-ion collisions are discussed. The inelastic cross sections, in particular for the process of the channel N$\Delta\rightarrow$NN, are parameterized in accordance with the available experimental data. It is found that the in-medium cross sections play a significant role on isospin emissions and result in a flat distribution for transverse flows and elliptic flows of free nucleons comparing with the in-vacuum ones. The rapidity distribution of the difference between neutron and proton transverse flows is sensitive to the stiffness of nuclear symmetry energy as a promising observable, which can not be influenced by the in-medium effect and collision centrality. Furthermore, the elliptic flow of free energetic nucleons in the case of the mass splitting of $m_{n}^{\ast}>m_{p}^{\ast}$ is also related to the symmetry energy. However, the pion flows weakly depend on the symmetry energy and the mass splitting.
\begin{description}
\item[PACS number(s)]
21.65.Ef, 24.10.Lx, 25.75.-q
\end{description}
\end{abstract}

\maketitle

\section{I. Introduction}

The high-density behavior of nuclear symmetry energy is of importance in understanding the properties of compact stars, in particular, the structure of neutron stars (mass and radius), the cooling of protoneutron stars, the nucleosynthesis during supernova explosion of massive stars etc \cite{St05, Kl06}. Up to now, the high-density information of the symmetry energy is poorly known \cite{Ba05,Li08,To10}. The importance of high-density symmetry energy in hadron-quark phase transition was stated and investigated thoroughly by Di Toro \emph{et al.} \cite{To11}. Heavy-ion collisions induced by neutron-rich beams at intermediate energies are an unique approach for extracting the information of isospin-asymmetric nuclear equation of state (EOS) in terrestrial laboratories. The symmetry energy in neutron-rich matter is usually expressed through the energy per nucleon as $E(\rho,\delta)=E(\rho,\delta=0)+E_{\textrm{sym}}(\rho)\delta^{2}+O(\delta^{2})$ in terms of baryon density $\rho=\rho_{n}+\rho_{p}$, relative neutron excess $\delta=(\rho_{n}-\rho_{p})/(\rho_{n}+\rho_{p})$, energy per nucleon in a symmetric nuclear matter $E(\rho,\delta=0)$ and nuclear symmetry energy per nucleon $E_{\textrm{sym}}$. From that, one can easily get the physical quantities of nuclear matter, such as pressure, chemical potential and single particle potential etc. Based on several complementary analysis of available experimental data associated with transport models, a symmetry energy of $E_{\textrm{sym}}(\rho)\approx 31.6(\rho/\rho_{0})^{\gamma}$ MeV with $\gamma=0.69-1.05$ was extracted for densities between 0.1$\rho_{0}$ and 1.2$\rho_{0}$ \cite{Li08}. However, predictions for the high-density symmetry energy based on various microscopical or phenomenological many-body theories diverge widely \cite{Di03,Ch05,Li06}. Moreover, the high-density behavior of nuclear symmetry energy predicted by transport models associated with the existing experimental data are also different largely because of the inconsistent treatment of the mean-field potentials and the unclear in-medium nuclear interaction and properties of resonances \cite{Fe06,Xi09,Fe10a,Ru11}, and an opposite conclusion was drawn. Furthermore, new experimental data related to high-density phase diagram and modifications of transport models including the in-medium effects in two-body collisions and in mean-field propagation, are very necessary.

Based on the QMD ('quantum' molecular dynamics) approach \cite{Ai87}, over the past decade the LQMD (Lanzhou quantum molecular dynamics) model has been successfully updated and applied to treat the nuclear dynamics from near Coulomb barrier energies up to relativistic energies (several \emph{A} GeV) \cite{Ch98,Fe05,Fe09,Fe10b}, in which the isospin effect, nuclear structure (shell effect) in mean-field potential, density dependence of symmetry energy, fermionic nature of nucleon in mean-field evolutions and in NN collisions, and inelastic channels including the in-medium effect etc, have been improved thoroughly. Furthermore, the momentum dependence of the symmetry potential was also included in the model, which results in an effective mass splitting of proton and neutron in nuclear medium and its influence on isospin emission was investigated \cite{Fe11a}. In this work, we present a review of the recent updated LQMD model. Collective flows of free nucleons (light clusters) and pions produced in heavy-ion collisions and influence of the in-medium effect such as NN cross section, optical potential in nuclear medium etc, collision centrality, mass splitting and stiffness of symmetry energy are investigated systematically. Extraction of the high-density symmetry energy from flow information are discussed.

The article is organized as follows. In Sec. II we give a description of the recent version of LQMD model. In-medium modification and its influence on isospin particle emissions, collective flows, and some promising observables for extracting the high-density symmetry energy are presented in Sec. III. In Sec. IV conclusions are discussed.

\section{II. Model description}

It is well known that the QMD model has the advantage of correctly treating the \emph{N}-body correlations caused by the overlapping of the Gaussian wave packets and by the event-by-event simulations. A similar Gaussian wave packet has been used in the stochastic mean-field (SMF) model and obtained progress in the description of nuclear multi-fragmentation reactions by Catania group \cite{Ba05,Ri05}. The same as in the QMD \cite{Ai87}, the wave function for each nucleon or resonance in the LQMD model is represented by a Gaussian wave packet. The width of the wave packet is fixed in the mean-field evolutions, but increases with the mass number of nuclide \cite{Fe05,Fe09,Fe10b}. The total \emph{N}-body wave function is assumed as the direct product of the coherent states and the anti-symmetrization is neglected. Therefore, the fermionic nature of nucleon is lost in the QMD-like models. We used a phase constraint approach to embody the nucleonic Fermi properties \cite{Fe05}. From the Wigner transformation for the total wave function, one can get the density distributions in coordinate and momentum space. In the LQMD model, the time evolutions of the baryons (nucleons and resonances ($\Delta$(1232), N*(1440), N*(1535))) and mesons in the system under the self-consistently generated mean-field are governed by Hamilton's equations of motion, which read as
\begin{eqnarray}
\dot{\mathbf{p}}_{i}=-\frac{\partial H}{\partial\mathbf{r}_{i}},
\quad \dot{\mathbf{r}}_{i}=\frac{\partial H}{\partial\mathbf{p}_{i}}.
\end{eqnarray}
Here we only consider the Coulomb interaction for charged hyperons. The Hamiltonian of baryons consists of the relativistic energy, the effective interaction potential and the momentum dependent interaction. The effective interaction potential is composed of the Coulomb interaction and the local potential
\begin{equation}
U_{int}=U_{Coul}+U_{loc}.
\end{equation}
The Coulomb interaction potential is written as
\begin{equation}
U_{Coul}=\frac{1}{2}\sum_{i,j,j\neq
i}\frac{e_{i}e_{j}}{r_{ij}}erf(r_{ij}/\sqrt{4L})
\end{equation}
where the $e_{j}$ is the charged number including protons and charged resonances. The $r_{ij}=|\mathbf{r}_{i}-\mathbf{r}_{j}|$ is the relative distance of two charged particles.

The local interaction potential is derived directly from the Skyrme energy-density functional and expressed as
\begin{equation}
U_{loc}=\int V_{loc}(\rho(\mathbf{r}))d\mathbf{r}.
\end{equation}
The local potential energy-density functional reads
\begin{eqnarray}
V_{loc}(\rho)=&& \frac{\alpha}{2}\frac{\rho^{2}}{\rho_{0}} + \frac{\beta}{1+\gamma}\frac{\rho^{1+\gamma}}{\rho_{0}^{\gamma}} +
g_{\tau}\rho^{8/3}/\rho_{0}^{5/3} + E_{sym}^{loc}(\rho)\rho\delta^{2}   \nonumber \\
&& + \frac{g_{sur}}{2\rho_{0}}(\nabla\rho)^{2} + \frac{g_{sur}^{iso}}{2\rho_{0}}[\nabla(\rho_{n}-\rho_{p})]^{2} ,
\end{eqnarray}
where the $\rho_{n}$, $\rho_{p}$ and $\rho=\rho_{n}+\rho_{p}$ are the neutron, proton and total densities, respectively, and the $\delta=(\rho_{n}-\rho_{p})/(\rho_{n}+\rho_{p})$ is the isospin asymmetry. The coefficients $\alpha$, $\beta$, $\gamma$, $g_{\tau}$, $g_{sur}$, $g_{sur}^{iso}$ are related to the Skyrme parameters $t_{0}, t_{1}, t_{2}, t_{3}$ and $x_{0}, x_{1}, x_{2}, x_{3}$ \cite{Fe05} and the parameter Sly6 is taken in the calculation with the values of 9.9 MeV, 22.9 MeV fm$^{2}$, -2.7 MeV fm$^{2}$ and 0.16 fm$^{-3}$ for $g_{\tau}$, $g_{sur}$, $g_{sur}^{iso}$ and $\rho_{0}$ respectively. The bulk parameters $\alpha$, $\beta$ and $\gamma$ are readjusted after inclusion the momentum term in order to reproduce the compression modulus of symmetric nuclear matter (here, K=230 MeV) and the binding energy of isospin symmetric nuclear matter at saturation density, which have the values of -296.6 MeV, 197 MeV and 1.143, respectively. The $E_{sym}^{loc}$ is the local part of the symmetry energy, which can be adjusted to mimic predictions of the symmetry energy calculated by microscopical or phenomenological many-body theories and has two-type forms as follows:
\begin{equation}
E_{sym}^{loc}(\rho)=\frac{1}{2}C_{sym}(\rho/\rho_{0})^{\gamma_{s}},
\end{equation}
and
\begin{equation}
E_{sym}^{loc}(\rho)=a_{sym}(\rho/\rho_{0})+b_{sym}(\rho/\rho_{0})^{2}.
\end{equation}
The parameters $C_{sym}$, $a_{sym}$ and $b_{sym}$ are taken as the values of 52.5 MeV, 43 MeV, -16.75 MeV and 23.52 MeV, 32.41 MeV, -20.65 MeV corresponding to the mass splittings of $m_{n}^{\ast}>m_{p}^{\ast}$ and $m_{n}^{\ast}<m_{p}^{\ast}$, respectively. The values of $\gamma_{s}$=0.5, 1., 2. have the soft, linear and hard symmetry energy with baryon density, respectively, and the Eq. (7) gives a supersoft symmetry energy, which cover the largely uncertain of nuclear symmetry energy, particularly at the supra-saturation densities. The local part of the symmetry energy can be adjusted to reflect the largely uncertain behavior of the symmetry energy at sub- and supra-normal densities.

A Skyrme-type momentum-dependent potential is used in the model as follows:
\begin{eqnarray}
U_{mom}=&& \frac{1}{2\rho_{0}}\sum_{i,j,j\neq i}\sum_{\tau,\tau'}C_{\tau,\tau'}\delta_{\tau,\tau_{i}}\delta_{\tau',\tau_{j}}\int\int\int
d\textbf{p}d\textbf{p}'d\textbf{r}  \nonumber \\
&& \times f_{i}(\textbf{r},\textbf{p},t)[\ln(\epsilon(\textbf{p}-\textbf{p}')^{2}+1)]^{2} f_{j}(\textbf{r},\textbf{p}',t).
\end{eqnarray}
The term is also given from the energy-density functional in nuclear matter as
\begin{eqnarray}
U_{mom}=&& \frac{1}{2\rho_{0}}\sum_{\tau,\tau'}C_{\tau,\tau'} \int\int\int d \textbf{p}d\textbf{p}'d\textbf{r}   \nonumber \\
&& \times f_{\tau}(\textbf{r},\textbf{p}) [\ln(\epsilon(\textbf{p}-\textbf{p}')^{2}+1)]^{2} f_{\tau'}(\textbf{r},\textbf{p}').
\end{eqnarray}
Here $C_{\tau,\tau}=C_{mom}(1+x)$, $C_{\tau,\tau'}=C_{mom}(1-x)$ ($\tau\neq\tau'$) and the isospin symbols $\tau$($\tau'$) represent proton or neutron. The sign of $x$ determines different mass splitting of proton and neutron in nuclear medium, e.g., positive signs corresponding to the case of $m^{\ast}_{n}<m^{\ast}_{p}$. The parameters $C_{mom}$ and $\epsilon$ was determined by fitting the real part of optical potential as a function of incident energy from the proton-nucleus elastic scattering data. In the calculation, we take the values of 1.76 MeV, 500 c$^{2}$/GeV$^{2}$ for the $C_{mom}$ and $\epsilon$, respectively, which result in the effective mass $m^{\ast}/m$=0.75 in nuclear medium at saturation density for symmetric nuclear matter. The parameter $x$ can be adjusted as the strength of the mass splitting, and the values of -0.65 and 0.65 are respective to the cases of $m^{\ast}_{n}>m^{\ast}_{p}$ and $m^{\ast}_{n}<m^{\ast}_{p}$. One can get the contribution of momentum-dependent interaction to symmetry energy for a cold nuclear with Fermi distribution for the phase-space density $f_{\tau}(\textbf{r},\textbf{p})=\rho_{\tau}(\textbf{r})\Theta(p_{F}(\tau)-|\textbf{p}|)/(4\pi p_{F}^{3}(\tau)/3)$ with the Fermi momentum $p_{F}(\tau)=\hbar(3\pi^{2}\rho_{\tau})^{1/3}$ \cite{Fe11a}. The symmetry energy per nucleon in the LQMD model is composed of three parts, namely the kinetic energy, the local part and the momentum dependence of the potential energy as
\begin{equation}
E_{sym}(\rho)=\frac{1}{3}\frac{\hbar^{2}}{2m}\left(\frac{3}{2}\pi^{2}\rho\right)^{2/3}+E_{sym}^{loc}(\rho)+E_{sym}^{mom}(\rho).
\end{equation}
After an expansion to second order around the normal density, the symmetry energy can be expressed as
\begin{equation}
E_{sym}(\rho) \approx E_{sym}(\rho_{0})+\frac{L}{3}\left(\frac{\rho-\rho_{0}}{\rho_{0}}\right)+\frac{K_{sym}}{18}\left(\frac{\rho-\rho_{0}}{\rho_{0}}\right)^{2}
\end{equation}
in terms of a slope parameter of $L=3\rho_{0}(\partial E_{sym}/\partial \rho)|_{\rho=\rho_{0}}$ and a curvature parameter of $K_{sym}=9\rho_{0}^{2}(\partial^{2} E_{sym}/\partial \rho^{2})|_{\rho=\rho_{0}}$. The values of slope parameters are 203.7 MeV, 124.9 MeV, 85.6 MeV and 74.7 MeV for the hard, linear, soft and supersoft symmetry energies, respectively, and the corresponding 448 MeV, -24.5 MeV, -83.5 MeV and -326 MeV for the curvature parameters. Figure 1 is a comparison of different stiffness of nuclear symmetry energy after inclusion of the momentum-dependent interactions. All cases cross at saturation density with the value of 31.5 MeV, which cover the largely uncertain of symmetry energy in a large density range.

\begin{figure}
\includegraphics[width=8 cm]{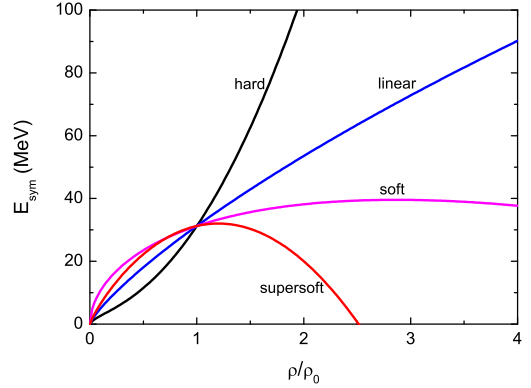}
\caption{\label{fig:epsart} (Color online) Density dependence of the nuclear symmetry energy for the cases of supersoft, soft, linear and hard trends with MDI interaction.}
\end{figure}

Combined Eq. (5) and Eq. (9), we get a density, isospin and momentum dependent single-nucleon potential in nuclear matter as follows:
\begin{eqnarray}
U_{\tau}(\rho,\delta,\textbf{p})=&&  \alpha\frac{\rho}{\rho_{0}}+\beta\frac{\rho^{\gamma}}{\rho_{0}^{\gamma}}+\frac{8}{3}g_{\tau}\rho^{5/3}/\rho_{0}^{5/3}+
E_{sym}^{loc}(\rho)\delta^{2}        \nonumber \\
&&  +  \frac{\partial E_{sym}^{loc}(\rho)}{\partial\rho}\rho\delta^{2} + E_{sym}^{loc}(\rho)\rho\frac{\partial\delta^{2}}{\partial\rho_{\tau}}   \nonumber \\
&&  + \frac{1}{\rho_{0}}C_{\tau,\tau} \int d\textbf{p}' f_{\tau}(\textbf{r},\textbf{p})[\ln(\epsilon(\textbf{p}-\textbf{p}')^{2}+1)]^{2}         \nonumber \\
&&  + \frac{1}{\rho_{0}}C_{\tau,\tau'} \int d\textbf{p}' f_{\tau'}(\textbf{r},\textbf{p})      \nonumber \\
&&  \times [\ln(\epsilon(\textbf{p}-\textbf{p}')^{2}+1)]^{2}.
\end{eqnarray}
Here $\tau\neq\tau'$, $\partial\delta^{2}/\partial\rho_{n}=4\delta\rho_{p}/\rho^{2}$ and $\partial\delta^{2}/\partial\rho_{p}=-4\delta\rho_{n}/\rho^{2}$. The nucleon effective (Landau) mass in nuclear matter of isospin asymmetry $\delta=(\rho_{n}-\rho_{p})/(\rho_{n}+\rho_{p})$ with $\rho_{n}$ and $\rho_{p}$ being the neutron and proton density, respectively, is calculated through the potential as $m_{\tau}^{\ast}=m_{\tau}/ \left(1+\frac{m_{\tau}}{|\textbf{p}|}|\frac{dU_{\tau}}{d\textbf{p}}|\right)$ with the free mass $m_{\tau}$ at Fermi momentum $\textbf{p}=\textbf{p}_{F}$. Therefore, the nucleon effective mass only depends on the momentum-dependent term of the nucleon optical potential. The isovector part of the optical potential, i.e., the symmetry or Lane potential \cite{La62}, can be evaluated from the expression $U_{sym}(\rho,\textbf{p})=(U_{n}(\rho,\delta,\textbf{p})-U_{p}(\rho,\delta,\textbf{p}))/2\delta$. Shown in Fig. 2 is a comparison of the momentum dependence of single-nucleon optical potential with the mass splittings of $m_{n}^{\ast}>m_{p}^{\ast}$ in the left window and $m_{n}^{\ast}<m_{p}^{\ast}$ in the right window for the hard and supersoft symmetry energies, respectively. One should note a cross appears in the case of $m_{n}^{\ast}>m_{p}^{\ast}$ at high momentum, but which does not take place in the mass splitting of $m_{n}^{\ast}<m_{p}^{\ast}$ and a broader separation exists with increasing the nucleon momentum. The symmetry potential is also being influenced by the mass splitting and just an opposite trend appears as shown in Fig. 3. A pronounced difference of both mass splittings can be constrained from elliptic flow in heavy-ion collisions, especially from the momentum (kinetic energy) distribution of $V_{2}^{n}-V_{2}^{p}$ at mid-rapidity \cite{Fe11b}. More specifically, the symmetry potential is plotted as functions of momentum and density in Fig. 4. Both the stiffness of symmetry energy and the mass splitting affect the spectrum. The effective mass is opposite in both mass splittings as shown in Fig. 5, which is pronounced in the 1\emph{A} GeV energies (1$\sim$3$\rho_{0}$). Therefore, both of the stiffness of nuclear symmetry energy and the mass splitting affect the reaction dynamics, in particular for the isospin emission in heavy-ion collisions.

\begin{figure*}
\includegraphics{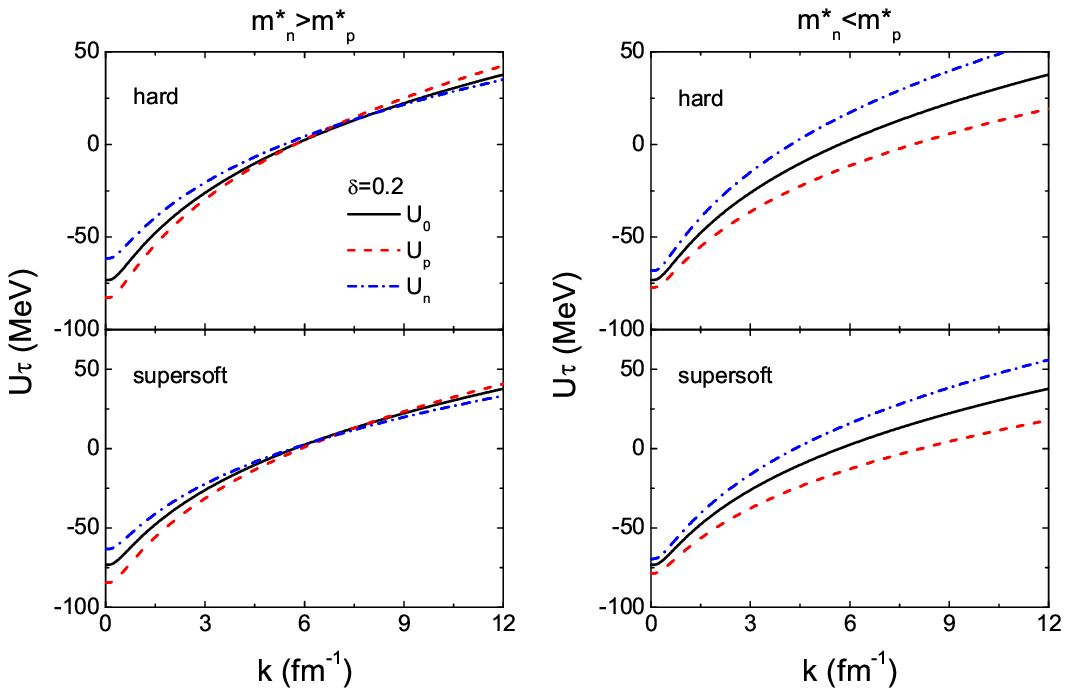}
\caption{\label{fig:wide} (Color online) Momentum dependence of single-nucleon optical potential for isospin symmetric matter ($\delta$=0) and neutron-rich matter ($\delta$=0.2) with the mass splittings of $m_{n}^{\ast}>m_{p}^{\ast}$ (left panel) and $m_{n}^{\ast}<m_{p}^{\ast}$ (right panel), respectively.}
\end{figure*}

\begin{figure*}
\includegraphics{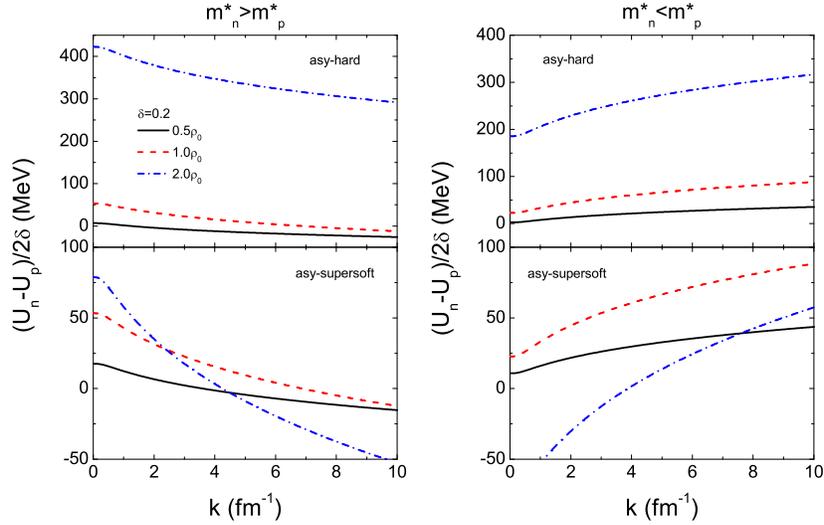}
\caption{\label{fig:wide} (Color online) Symmetry potential as a function of momentum with the hard and supersoft symmetry energies and with different mass splittings at nuclear matter density 0.5$\rho_{0}$, $\rho_{0}$ and 2$\rho_{0}$, respectively.}
\end{figure*}

\begin{figure*}
\includegraphics{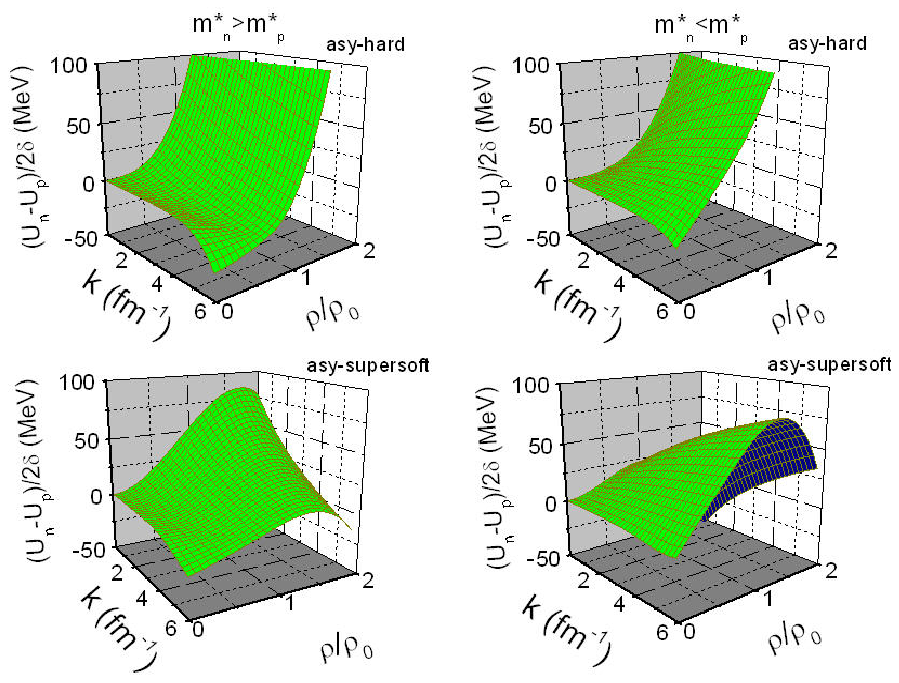}
\caption{\label{fig:wide} (Color online) Contour of symmetry potential versus momentum and density of nuclear matter with different symmetry energies and mass splittings.}
\end{figure*}

\begin{figure*}
\includegraphics{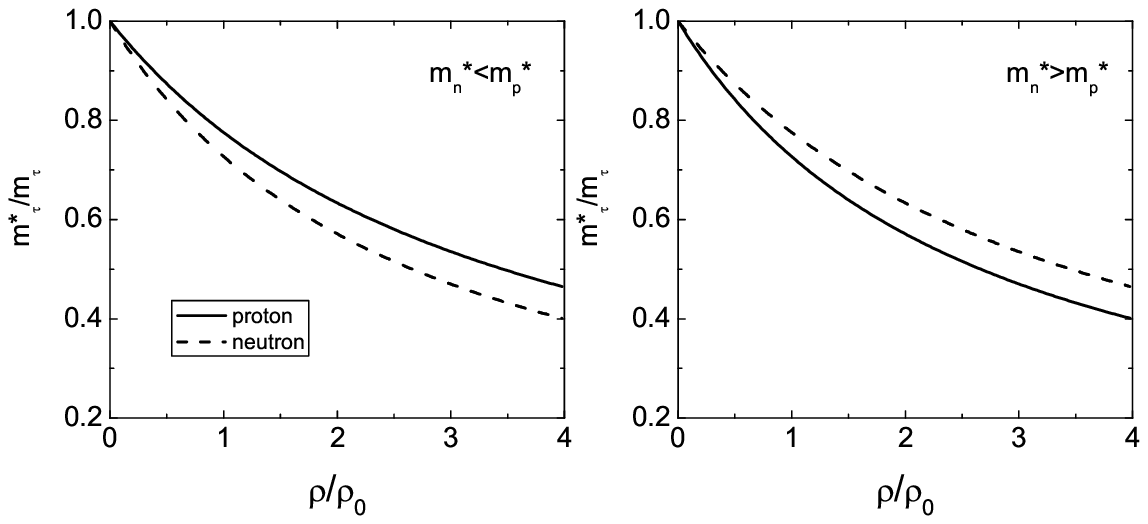}
\caption{\label{fig:wide} Effective mass of proton and neutron in nuclear medium as a function of baryon density with the isospin asymmetry of $\delta$=0.2 for the mass splittings of $m_{n}^{\ast}>m_{p}^{\ast}$ (left panel) and $m_{n}^{\ast}<m_{p}^{\ast}$ (right panel), respectively.}
\end{figure*}

The same with the QMD or BUU (Boltzmann-Uehling-Uhlenbeck) model, a hard core scattering in two-particle collisions is assumed by using Monte Carlo procedures, in which the scattering of two particles is determined by a geometrical minimum distance criterion $d\leq\sqrt{0.1\sigma_{tot}/\pi}$ fm weighted by the Pauli blocking of the final states \cite{Ai87,Be88}. Here, the total NN cross section $\sigma_{tot}$ in mb is the sum of the elastic and all inelastic cross sections. The probability reaching a channel in a collision is calculated by its contribution of the channel cross section to the total cross section as $P_{ch}=\sigma_{ch}/\sigma_{tot}$. The choice of the channel is done randomly by the weight of the probability. We parametrized the total, elastic and inelastic NN cross sections in accordance with the available experimental data in free space as shown in Fig. 6. One expects the in-medium NN elastic collisions are reduced in comparison with the free-space ones. The in-medium elastic cross section is scaled according to the effective mass that was used in the BUU model \cite{De02}. Following the IBUU04 model \cite{Li05}, the elastic cross section in the nuclear medium is evaluated through $\sigma_{NN}^{medium}=(\mu^{\ast}_{NN}/\mu_{NN})^{2}\sigma_{NN}^{free}$ with the $\mu^{\ast}_{NN}$ and $\mu_{NN}$ being the reduced masses of colliding nucleon pairs in the medium and in the free space, respectively. Shown in Fig. 7 is a comparison of the scaling factor as functions of baryon density, nucleon momentum and isospin asymmetry for the different mass splittings of $m_{n}^{\ast}>m_{p}^{\ast}$ and $m_{n}^{\ast}<m_{p}^{\ast}$, respectively. It is interesting to notice that a splitting of $nn$ (neutron-neutron), $pn$ (proton-neutron) and $pp$ (proton-proton) exists because of the mixing of neutron and proton effective masses. The mass splitting results in an opposite splitting of the in-medium cross sections of $nn$ and $pp$ channels. Recent microscopic calculations from the Brueckner-Hartree-Fock and the relativistic Brueckner-Hartree-Fock basically favor the variation of the in-medium cross section with baryon density, nucleon momentum and isospin asymmetry \cite{Zh07,Fu01}.

\begin{figure*}
\includegraphics{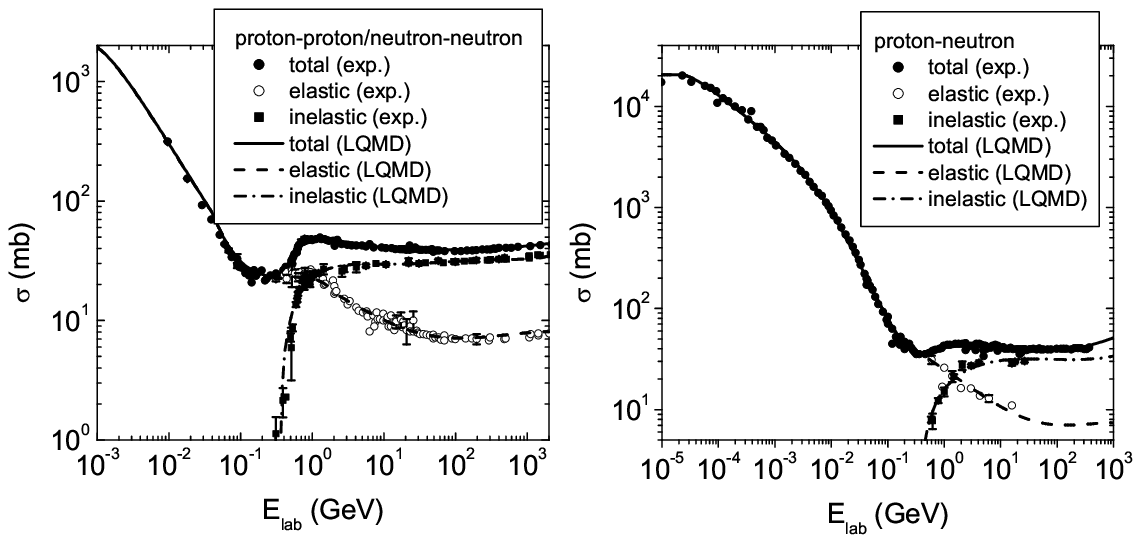}
\caption{\label{fig:wide} Nucleon-nucleon cross sections of total, elastic and inelastic channels in free space parameterized in the LQMD model as a function of nucleon incident energy in the laboratory frame.}
\end{figure*}

\begin{figure*}
\includegraphics{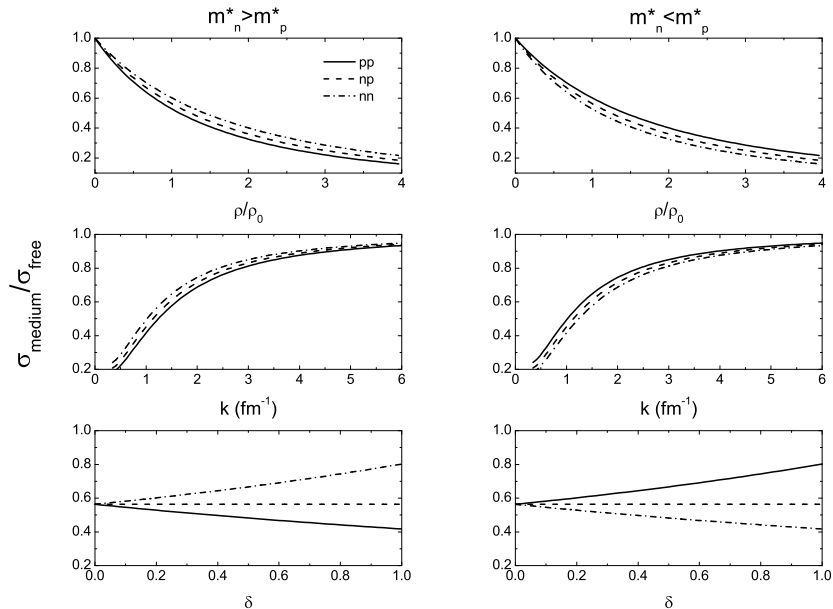}
\caption{\label{fig:wide} In-medium nucleon-nucleon elastic cross section scaled by nucleon effective mass as functions of baryon density ($\delta=0.2$), nucleon momentum ($\delta=0.2$) and isospin asymmetry ($\rho=\rho_{0}$).}
\end{figure*}

\begin{figure}
\includegraphics[width=8 cm]{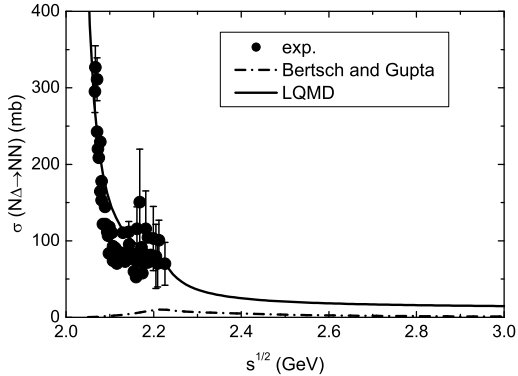}
\caption{\label{fig:epsart} Comparison of the cross section of the channel N$\Delta\rightarrow$NN parameterized in the model and compared with the approach by Bertsch and Gupta used in the BUU model and with the experimental data \cite{Ho96}.}
\end{figure}

The primary products in nucleon-nucleon (NN) collisions in the region of 1\emph{A} GeV energies are the resonances $\Delta$(1232), $N^{\ast}$(1440), $N^{\ast}$(1535) and the pions. We have included the reaction channels as follows:
\begin{eqnarray}
&& NN \leftrightarrow N\triangle, \quad  NN \leftrightarrow NN^{\ast}, \quad  NN
\leftrightarrow \triangle\triangle,  \nonumber \\
&& \Delta \leftrightarrow N\pi,  N^{\ast} \leftrightarrow N\pi,  NN \rightarrow NN\pi (s-state),  \nonumber \\
&& N^{\ast}(1535) \rightarrow N\eta.
\end{eqnarray}
At the considered energies, there are mostly $\Delta$ resonances which disintegrate into a $\pi$ and a nucleon in the evolutions. However, the $N^{\ast}$ yet gives considerable contribution to the energetic pion yields. The energy and momentum-dependent decay widths are used in the model for the $\Delta$(1232) and $N^{\ast}$(1440) resonances \cite{Fe09}. We have taken a constant width of $\Gamma$=150 MeV for the $N^{\ast}$(1535) decay. The in-medium effect of inelastic cross section is of importance in the estimation of meson production, which has been an interesting subject and still is an open problem because of the coupling of each channels \cite{La01,Ga05,Pr07}. A parameterized cross section is used in the LQMD model for the channel of N$\Delta\rightarrow$NN by fitting the available experimental data \cite{Ho96} with a formula by Wolf \emph{et al.} \cite{Wo93}, and compared with the cross section from the NN$\rightarrow$N$\Delta$ channel using the detailed balancing principle as shown in Fig. 8. The in-medium modifications of the absorption of $\Delta$ in heavy-ion collisions are also investigated in Refs. \cite{Da91,Li06b}. Calculations based on relativistic Dirac-Brueckner also favor the decrease trend of the $\Delta$ absorption cross section with increasing energy in nuclear medium at low densities (less than 2$\rho_{0}$) \cite{Ha87}. We used the parametrized cross sections calculated by the one-boson exchange model \cite{Hu94} for resonance production and the absorption of $N^{\ast}$ with the detailed balancing principle. The strangeness is created by inelastic hadron-hadron collisions. We included the channels as follows:
\begin{eqnarray}
&& BB \rightarrow BYK,  BB \rightarrow BBK\overline{K},  B\pi \rightarrow YK,  B\pi \rightarrow NK\overline{K}, \nonumber \\
&& Y\pi \rightarrow B\overline{K}, \quad  B\overline{K} \rightarrow Y\pi, \quad YN \rightarrow \overline{K}NN.
\end{eqnarray}
Here the B strands for (N, $\triangle$, N$^{\ast}$) and Y($\Lambda$, $\Sigma$), K(K$^{0}$, K$^{+}$) and $\overline{K}$($\overline{K^{0}}$, K$^{-}$). The elastic scattering between strangeness and baryons are considered through the channels $KB \rightarrow KB$, $YB \rightarrow YB$ and $\overline{K}B \rightarrow \overline{K}B$. At the moment, we do not implement the charge-exchange reactions between the $KN \rightarrow KN$ and $YN \rightarrow YN$ channels, such as $K^{0}p\rightarrow K^{+}n$, $K^{+}n\rightarrow K^{0}p$, $\Lambda n\rightarrow \Sigma^{-}p$, $\Lambda n\rightarrow \Sigma^{0}n$ etc, which would be important in evaluating the isospin ratios of $K^{0}/K^{+}$ and $\Sigma^{-}/\Sigma^{+}$ for probing the high-density symmetry energy.

Analogously to baryons, the evolution of mesons (here mainly pions and kaons) is also determined by the Hamiltonian, which is given by
\begin{eqnarray}
H_{M}&& = \sum_{i=1}^{N_{M}}\left( V_{i}^{\textrm{Coul}} + \omega(\textbf{p}_{i},\rho_{i}) \right).
\end{eqnarray}
Here the Coulomb interaction is given by
\begin{equation}
V_{i}^{\textrm{Coul}}=\sum_{j=1}^{N_{B}}\frac{e_{i}e_{j}}{r_{ij}},
\end{equation}
where the $N_{M}$ and $N_{B}$ are the total numbers of mesons and baryons including charged resonances. The pion optical potential $\textrm{Re}V_{\pi}^{opt}$ originates
from the medium effects in the hot and dense nuclear matter. In the calculation, we can also choose the value in vacuum, that is, the $\textrm{Re}V_{\pi}^{opt}$ is set equal zero. The influence of the pionic mean field in heavy-ion collisions on the transverse momentum distribution was investigated by using a phenomenological ansatz and a microscopic approach based on the $\Delta$-hole model by Fuchs \emph{et al.} \cite{Fu97}. Here we use the phenomenological ansatz suggested by Gale and Kapusta \cite{Ga87}. Then the dispersion relation reads
\begin{eqnarray}
\omega(\textbf{p}_{i},\rho_{i})=\sqrt{(|\textbf{p}_{i}|-p_{0})^{2}+m_{0}^{2}}-U,  \\
U=\sqrt{p_{0}^{2}+m_{0}^{2}}-m_{\pi},  \\
m_{0}=m_{\pi}+6.5(1-x^{10})m_{\pi},   \\
p_{0}^{2}=(1-x)^{2}m_{\pi}^{2}+2m_{0}m_{\pi}(1-x).
\end{eqnarray}
The phenomenological medium dependence on the baryon density is introduced via the coefficient $x(\rho_{i})=\exp(-a(\rho_{i}/\rho_{0}))$ with the parameter
$a=0.154$ and the saturation density $\rho_{0}$ in nuclear matter. Influence of the in-medium effects on the charged pion ratio is also investigated in Ref. \cite{Xu10}. We consider two scenarios for kaon (antikaon) propagation in nuclear medium, one with and one without medium modification. From the chiral Lagrangian the kaon and antikaon energy in the nuclear medium can be written as \cite{Ka86,Li97}
\begin{equation}
\omega(\textbf{p}_{i},\rho_{i})=\left[m_{K}^{2}+\textbf{p}_{i}^{2}-a_{K}\rho_{i}^{S}+
(b_{K}\rho_{i})^{2}\right]^{1/2}+b_{K}\rho_{i},
\end{equation}
\begin{equation}
\omega(\textbf{p}_{i},\rho_{i})=\left[m_{\overline{K}}^{2}+\textbf{p}_{i}^{2}-a_{\overline{K}}\rho_{i}^{S}+
(b_{K}\rho_{i})^{2}\right]^{1/2}-b_{K}\rho_{i},
\end{equation}
respectively. Here the $b_{K}=3/(8f_{\pi}^{2})\approx$0.32 GeVfm$^{3}$, the $a_{K}$ and $a_{\overline{K}}$ are 0.18 GeV$^{2}$fm$^{3}$ and 0.3 GeV$^{2}$fm$^{3}$, respectively, which result in the strength of repulsive kaon-nucleon potential and of attractive antikaon-nucleon potential with the values of 25.5 MeV and -96.8 MeV at saturation baryon density. Shown in Fig. 9 is comparison of the pion self-energy and optical potential as a function of momentum. One notices that an attractive potential is used in the LQMD model, in particular for energetic pions in nuclear medium. We calculated the pion emissions after inclusion of all, part and none of the optical potential and Coulomb interaction of pions with charged baryons in the reaction $^{124}$Sn+$^{124}$Sn for near-central collisions (b=1 fm) at incident energy of E$_{lab}$=400\emph{A} MeV as shown in Fig. 10. Without the Coulomb and pion potential, a flat $\pi^{-}/\pi^{+}$ ratio appears with the transverse momentum. The Coulomb force enhances the energetic $\pi^{+}$ production because of its repulsive force with charged baryons, which is opposite for $\pi^{-}$ emission, consequently increases the low-energy $\pi^{-}/\pi^{+}$ ratio and reduces the high-energy yields. However, the pion optical potential is always attractive for both charged pions. Calculations in Ref. \cite{Fe10b} show that the pion potential slightly affects the total $\pi^{-}/\pi^{+}$ yields owing to without distinguishing isospin effects for the charged pion potentials. Experimental measurements on the distributions of the transverse momentum or kinetic energy for the $\pi^{-}/\pi^{+}$ ratio will be helpful in understanding the pion optical potential in nuclear medium.

\begin{figure*}
\includegraphics{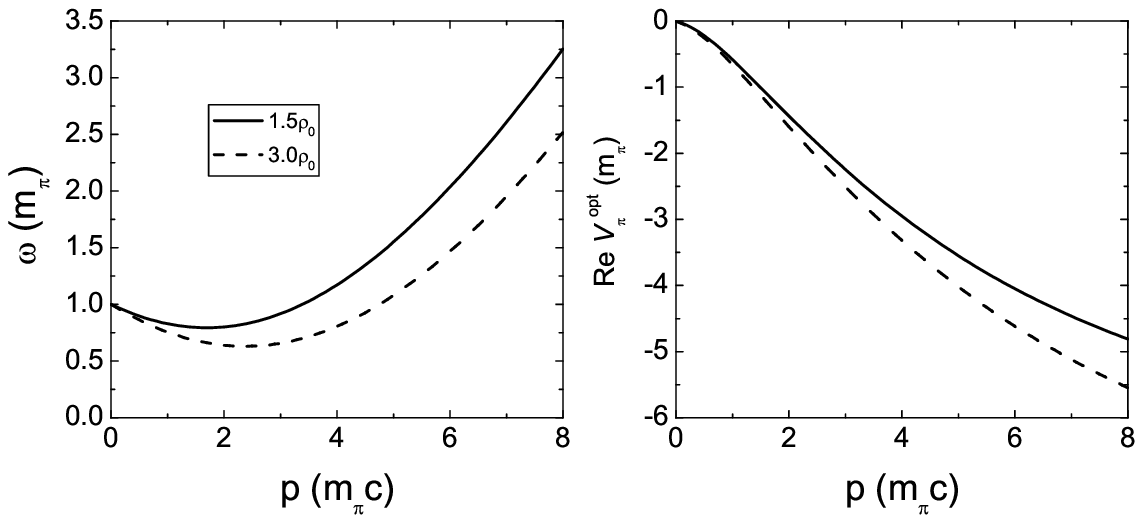}
\caption{\label{fig:wide} Momentum dependence of pion self-energy and optical potential at the baryon densities of 1.5$\rho_{0}$ and 3$\rho_{0}$, respectively.}
\end{figure*}

\begin{figure}
\includegraphics[width=8 cm]{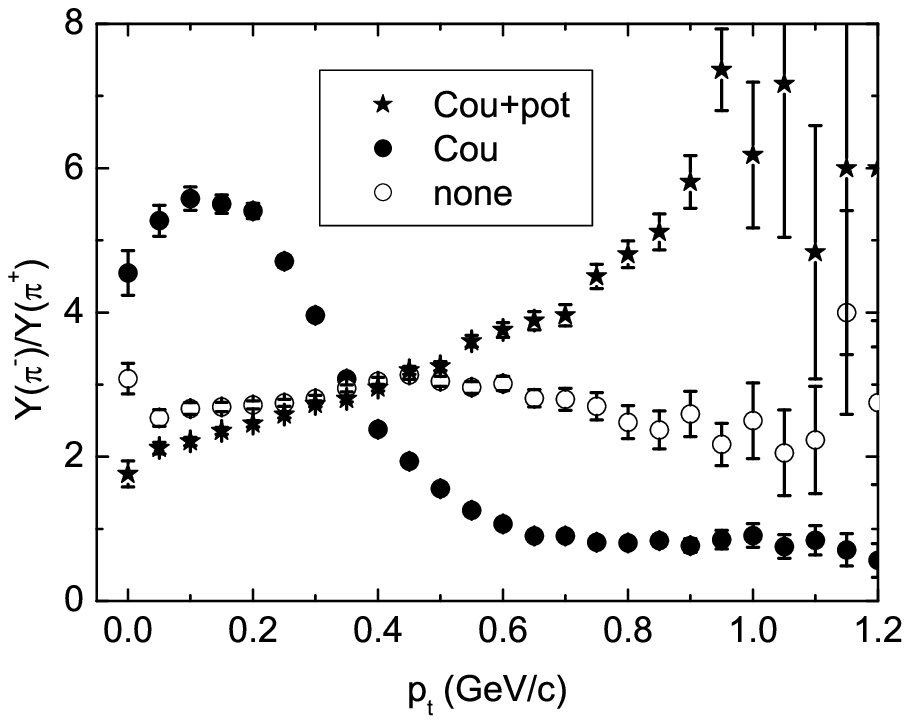}
\caption{\label{fig:epsart} Transverse momentum distributions of the ratios of $\pi^{-}$ to $\pi^{+}$ yields with including all, only Coulomb potential and none of the optical potential and Coulomb force of pions with baryons in the reaction $^{124}$Sn+$^{124}$Sn at incident energy of E$_{lab}$=400\emph{A} MeV.}
\end{figure}

\section{III. Results and discussions}

Collective flows in heavy-ion collisions have been studied well to extract the information of EOS at extreme conditions, i.e., high density, high temperature and large isospin asymmetry etc., and to explore the knowledge of in-medium properties and nuclear dynamics, such as nucleonic flow, light cluster flow, and meson flow etc \cite{Ru11,Ra99,Re07}. It is possible to reconstruct the reaction plane with flow analysis and hence to study azimuthal correlations of emitted particles. The flow information can be expressed as the first and second coefficients from the Fourier expansion of the azimuthal distribution $\frac{dN}{d\phi}(y,p_{t})=N_{0}(1+2V_{1}(y,p_{t})\cos(\phi)+2V_{2}(y,p_{t})\cos(2\phi))$ \cite{Ol92}, where the $p_{t}=\sqrt{p_{x}^{2}+p_{y}^{2}}$ and $y$ are the transverse momentum and the longitudinal rapidity along the beam direction, respectively. The directed (transverse) flow is defined as the first coefficient and expressed as $V_{1}=\langle p_{x}/p_{t} \rangle$, which provides the information of the azimuthal anisotropy of the transverse emission. The elliptic flow $V_{2}=\langle (p_{x}/p_{t})^{2}-(p_{y}/p_{t})^{2} \rangle$ gives the competition between the in-plane ($V_{2}>$0) and out-of-plane ($V_{2}<$0, squeeze out) emissions. The brackets indicate averaging over all events in accordance with a specific class such as rapidity or transverse momentum cut. Shown in Fig. 11 is a comparison of rapidity distributions of transverse and elliptic flows of free nucleons for near-central (b=1 fm) and semi-central (b=6 fm) $^{197}$Au+$^{197}$Au collisions at the incident energy of 400\emph{A} MeV with a choice of the mass splitting of $m_{n}^{\ast}>m_{p}^{\ast}$ for the hard and supersoft symmetry energies, respectively. One notices that the difference of neutron and proton transverse flows is small for the near-central collisions, but pronounced for the semi-central collisions, in particular around projectile rapidity. The elliptic flow disappears for the near-central heavy-ion collisions and the influence of symmetry energy on the elliptic flow distribution is very weak. The difference of neutron and proton elliptic flows is obvious in the semi-central collisions. In this work, the free nucleons and fragments are constructed with a coalescence model at freeze-out, in which nucleons of the reaction system are considered to belong to one cluster in the phase space with the relative momentum smaller than $P_{0}$ and with the relative distance less than $R_{0}$ (here, $P_{0}$=200 MeV/c and $R_{0}$=3 fm). The set of the parameters can reproduce the distribution of experimental charged fragments in a large region of energies \cite{Fe10b}.

\begin{figure*}
\includegraphics{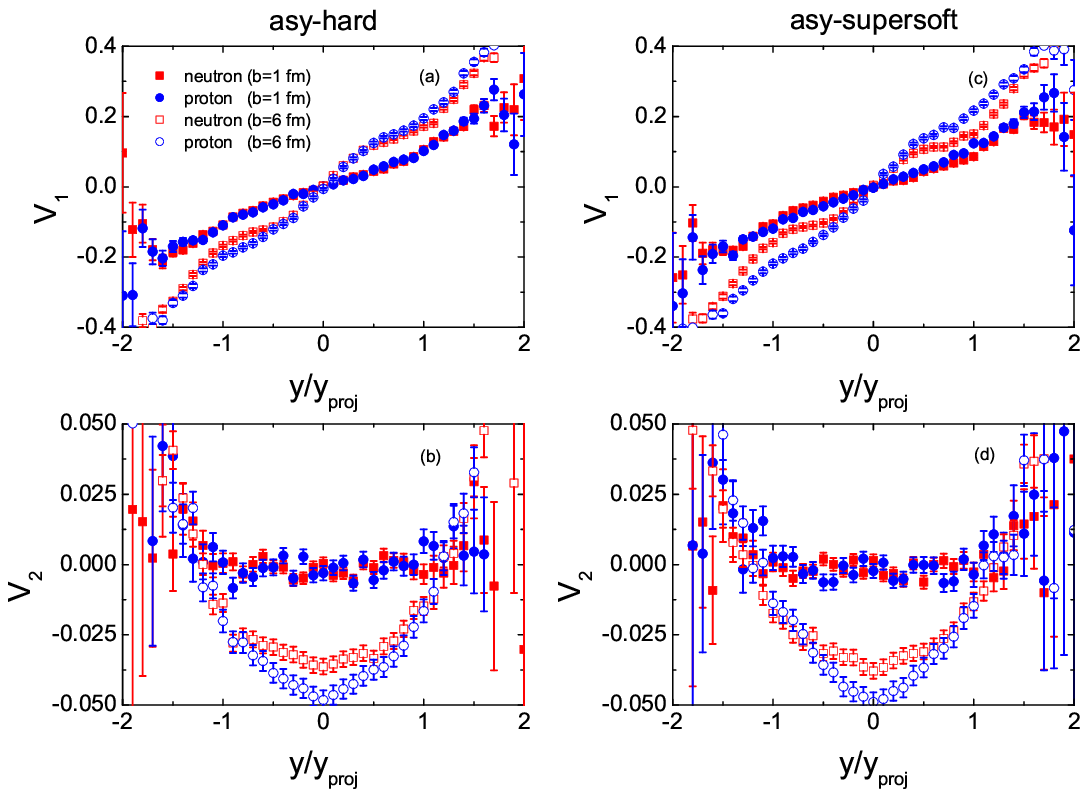}
\caption{\label{fig:wide} (Color online) Rapidity distributions of directed and elliptic flows of free nucleons in the $^{197}$Au+$^{197}$Au reaction at the incident energy of 400 MeV/nucleon for the near central (b=1 fm) and semi-central (b=6 fm) collisions using the NN elastic cross section in free space with the mass splitting of $m_{n}^{\ast}>m_{p}^{\ast}$ for the hard (left panel) and supersoft (right panel) symmetry energies, respectively.}
\end{figure*}

It has been well known that the transverse flow sensitively depends on the variation of in-medium NN cross sections, especially around the balance energies \cite{Xu91,Li93}. As a comparison, we calculated the transverse flow of free nucleons as shown in Fig. 12 for the $^{124}$Sn+$^{124}$Sn reaction at the incident energy of 400 MeV/nucleon using the elastic cross sections in free space and in nuclear medium, respectively. One notes that the in-medium cross section leads to a flat distribution for the near-central collisions. However, its influence on the semi-central collisions is slightly. The in-medium cross section also has a strongly influence on the elliptic flow distribution as shown in Fig. 13, which reduces the out-of-plane emissions of free nucleons, especially in the domain of mid-rapidity. Shown in Fig. 14 is the in-medium effect on light clusters with the hard (left window) and supersoft (right window) symmetry energies. It is seen that the reduced in-medium cross sections slightly modify the collective flows of light clusters ($^{3}$H and $^{3}$He).

\begin{figure*}
\includegraphics{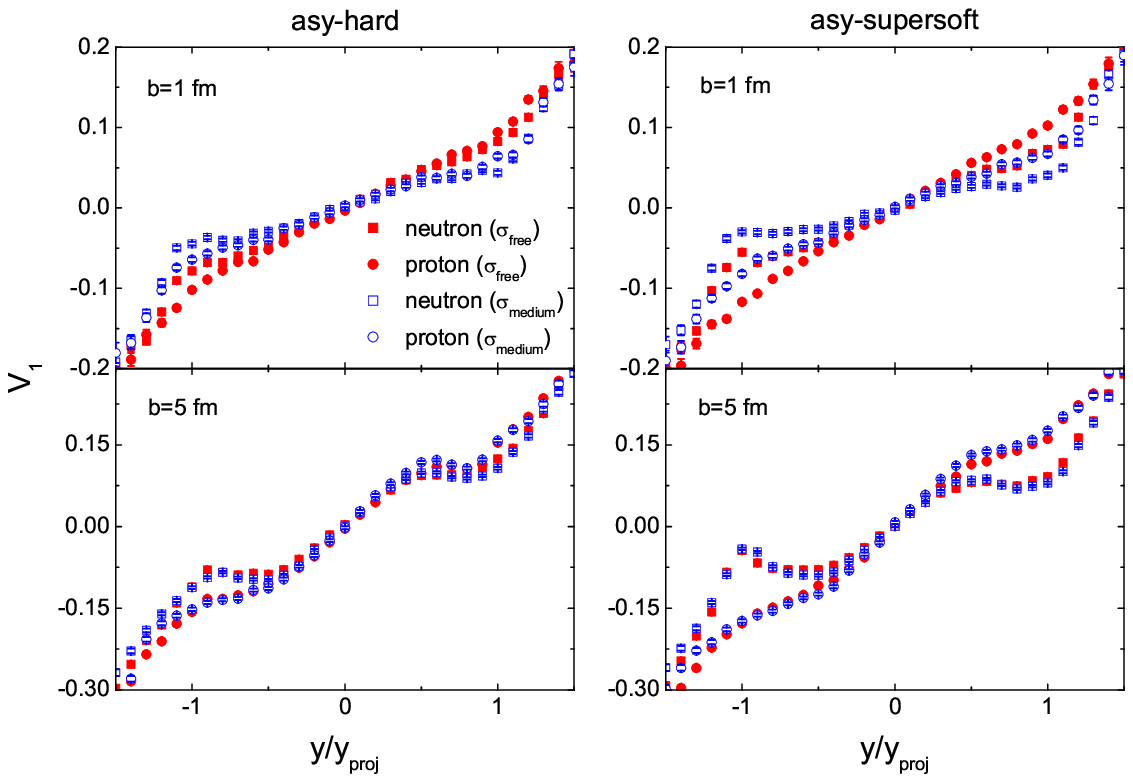}
\caption{\label{fig:wide} (Color online) Comparison of the transverse flow of free nucleons with in-vacuum and in-medium cross sections in the $^{124}$Sn+$^{124}$Sn reaction at the incident energy of 400 MeV/nucleon for the near-central (b=1 fm) and semi-central (b=6 fm) collisions with the mass splitting of $m_{n}^{\ast}>m_{p}^{\ast}$ for the hard (left panel) and supersoft (right panel) symmetry energies, respectively.}
\end{figure*}

\begin{figure*}
\includegraphics{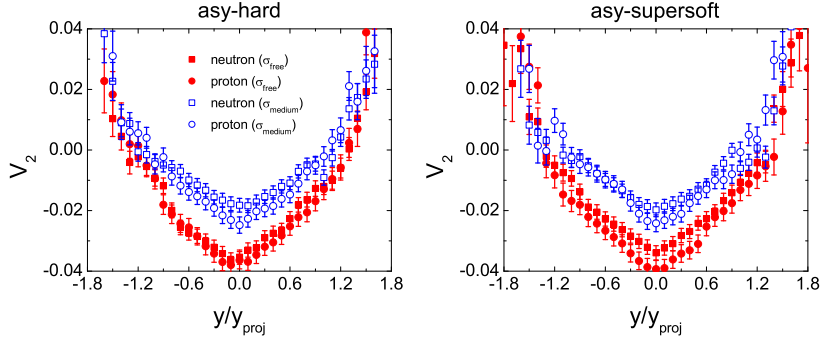}
\caption{\label{fig:wide} (Color online) The same as in Fig. 12, but for the elliptic flow for semi-central collisions (b=5 fm).}
\end{figure*}

\begin{figure*}
\includegraphics{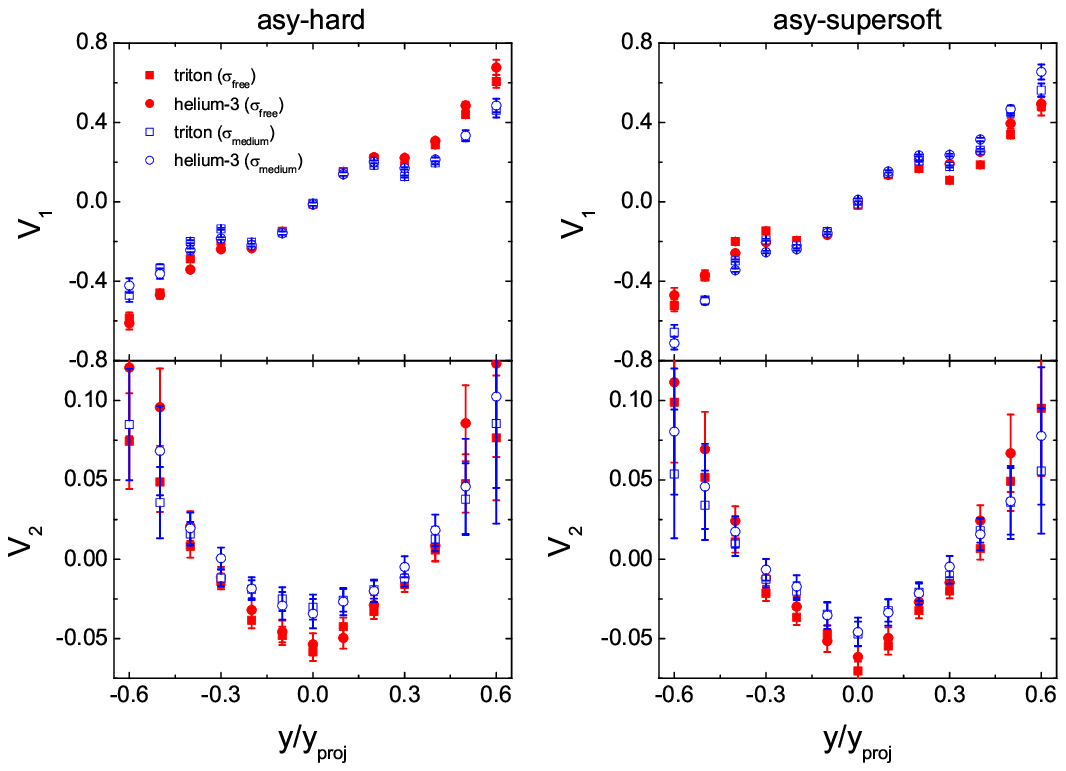}
\caption{\label{fig:wide} (Color online) Rapidity distributions of directed and elliptic flows of light clusters in the $^{124}$Sn+$^{124}$Sn reaction at a beam energy of 400\emph{A} MeV for the semi-central (b=5 fm) collisions with the mass splitting of $m_{n}^{\ast}>m_{p}^{\ast}$ for the hard (left panel) and supersoft (right panel) symmetry energies, respectively.}
\end{figure*}

In order to extract the density dependence of symmetry energy, especially for the high-density information, we calculated the flow difference of free nucleons as a function longitudinal rapidity as shown in Fig. 15. One can see that the spectrum is sensitive to the stiffness of symmetry energy in both near-central and semi-central collisions, especially around the projectile rapidity. The NN cross sections in free space and in nuclear medium basically does not change the distributions. A supersoft symmetry energy has a more pronounced spectrum. As a comparison, we also present the transverse flow difference of $^{3}$H and $^{3}$He as shown in Fig. 16. A similar structure appears, but weakly depends on the symmetry energy. A narrow rapidity distribution is observed for the light cluster emission because a number of free nucleons already take away much kinetic energies dissipated from the reaction zone of colliding partners. Shown in Fig. 17 is a comparison of transverse momentum distribution of elliptic flow of free nucleons with the hard and supersoft symmetry energies for the mass splittings of $m_{n}^{\ast}>m_{p}^{\ast}$ in the left panel and of $m_{n}^{\ast}<m_{p}^{\ast}$ in the right panel, respectively. The influence of symmetry energy on the elliptic flow is obvious only in the mass splitting of $m_{n}^{\ast}>m_{p}^{\ast}$, in particular at high transverse momentum. It should be mentioned that the distribution of elliptic flow is sensitively dependent on the mass splitting \cite{Fe11b}. Constraint of the high-density symmetry energy from elliptic flow has been performed in experiment \cite{Ru11} and more precise measurements are planned at RIKEN and GSI.

\begin{figure*}
\includegraphics{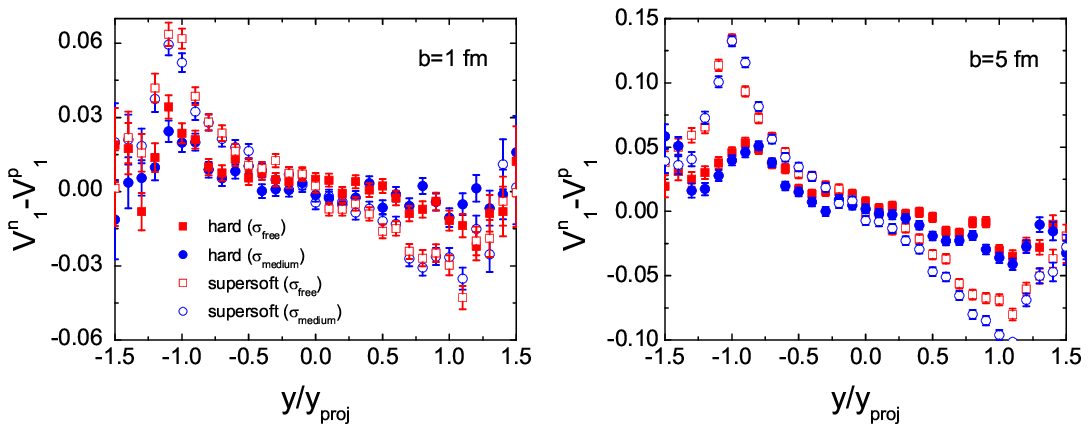}
\caption{\label{fig:wide} (Color online) The difference between neutron and proton directed flows in the $^{124}$Sn+$^{124}$Sn reaction at the energy of 400\emph{A} MeV .}
\end{figure*}

\begin{figure}
\includegraphics[width=8 cm]{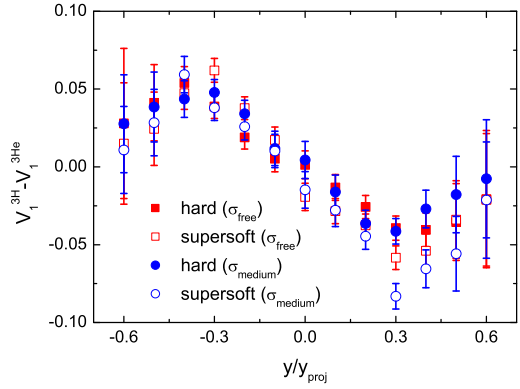}
\caption{\label{fig:epsart} (Color online) The same as in Fig. 15, but for the flow difference of $^{3}$H and $^{3}$He for semi-central collisions (b=5 fm).}
\end{figure}

\begin{figure*}
\includegraphics{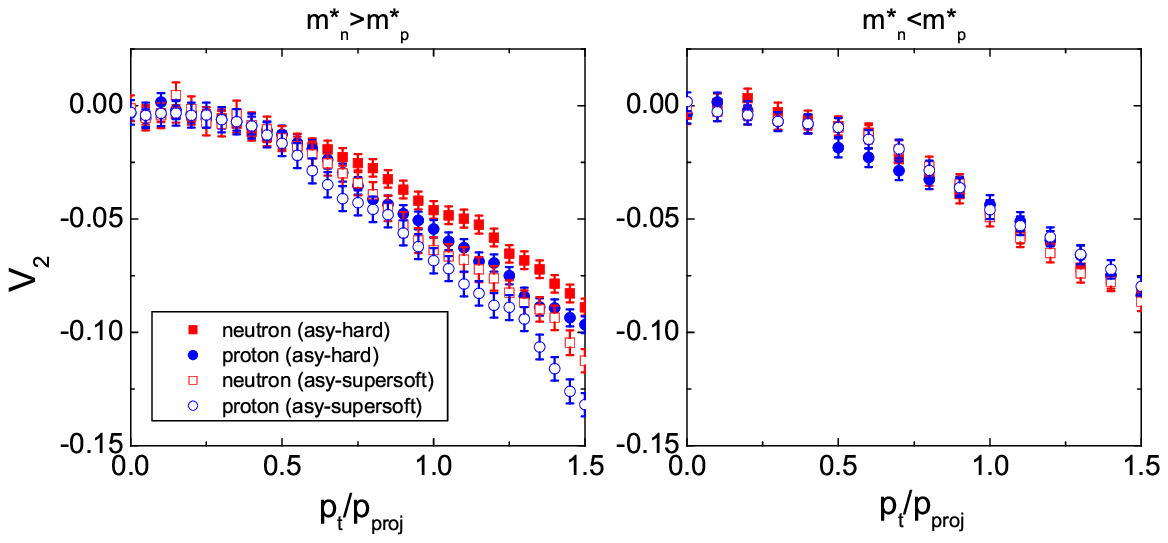}
\caption{\label{fig:wide} (Color online) The same as in Fig. 15, but for the momentum dependence of elliptic flow for semi-central collisions (b=5 fm) within the rapidity bin $|y/y_{proj}|<$0.3 for the mass splittings of $m_{n}^{\ast}>m_{p}^{\ast}$ (left panel) and of $m_{n}^{\ast}<m_{p}^{\ast}$ (right panel), respectively.}
\end{figure*}

Production of pions in heavy-ion collisions at near-threshold energies has been verified as a sensitive probe of nuclear symmetry energy, in particular the $\pi^{-}/\pi^{+}$ excitation functions \cite{Fe06,Xi09,Fe10a}, which can be easily detected in experimentally and be mainly created in the domain at supra-saturation densities of compressed nuclear matter. Shown in Fig. 18 is a comparison of transverse flows of $\pi^{-}$ and $\pi^{+}$ in $^{197}$Au and $^{197}$Au collisions at a near-threshold energy of 400 MeV/nucleon with the hard and supersoft symmetry energies for different mass splittings of $m_{n}^{\ast}>m_{p}^{\ast}$ in the left window and of $m_{n}^{\ast}<m_{p}^{\ast}$ in the right window, respectively. It can be seen that the symmetry energy and mass splitting basically does not affect the flow structure. Furthermore, an antiflow of $\pi^{+}$ (solid and blank squares) is observed in the both mass splittings by comparing with the transverse flow of free nucleons in Fig. 11. The phenomena is caused by the fact that pions produced in the 'fire ball' are absorbed again by participant nucleons (shadowing effect). A more pronounced flow ($\pi^{-}$) or antiflow ($\pi^{+}$) is observed comparing with the collisions at a higher incident energy of 1.5 GeV/nucleon \cite{Fe10b}. The antiflow of $\pi^{+}$ in heavy-ion collisions was also investigated using BUU \cite{Li94} and IQMD model \cite{Ba95}. For a comparison, we also calculated the flow difference between $\pi^{+}$ and $\pi^{-}$ with different symmetry energy and mass splitting as shown in Fig. 19. Although a pronounced flow difference ($V_{1}^{\pi^{+}}-V_{1}^{\pi^{-}}$) appears around the projectile rapidity, the influence of symmetry energy and mass splitting on the spectrum is still very weak.

\begin{figure*}
\includegraphics{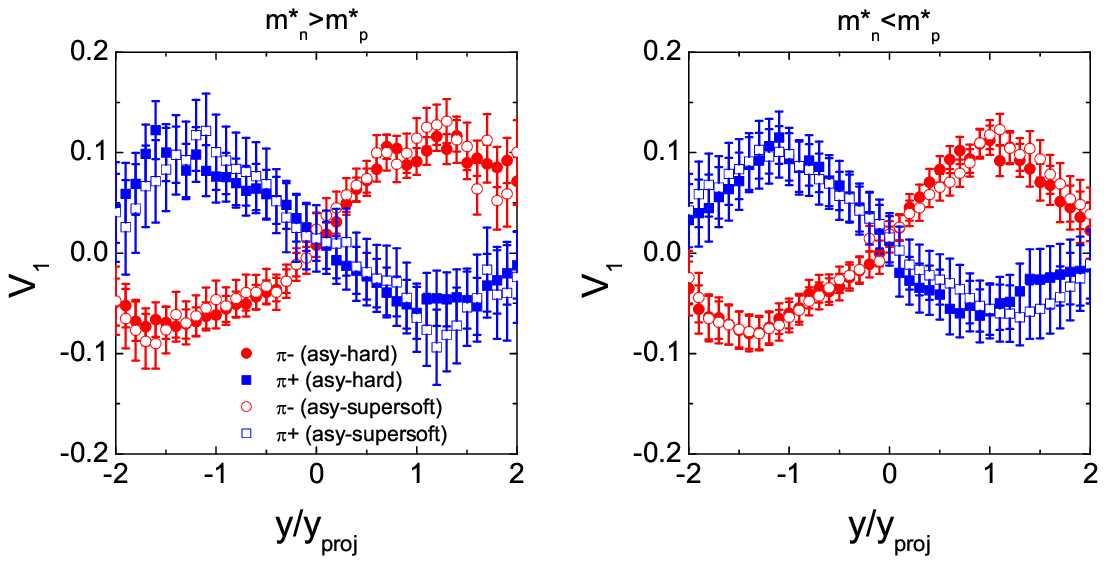}
\caption{\label{fig:wide} (Color online) Rapidity distributions of directed flows of charged pions in the $^{197}$Au+$^{197}$Au reaction at the incident energy of 400\emph{A} MeV for the semi-central (b=6 fm) collisions with the mass splittings of $m_{n}^{\ast}>m_{p}^{\ast}$ (left panel) and of $m_{n}^{\ast}<m_{p}^{\ast}$ (right panel), respectively.}
\end{figure*}

\begin{figure*}
\includegraphics{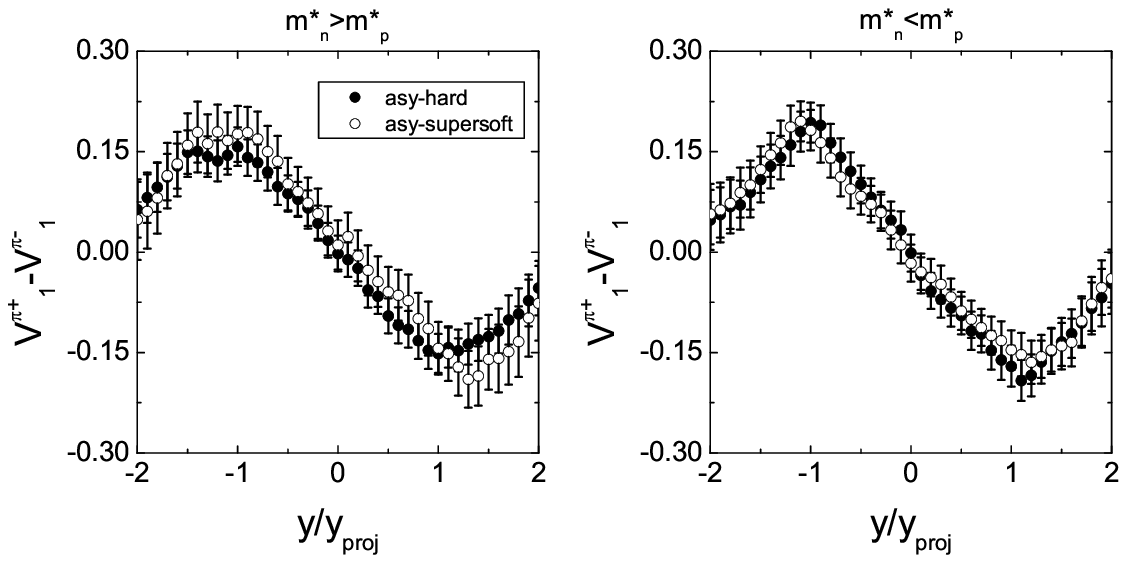}
\caption{\label{fig:wide} The directed flow difference between $\pi^{+}$ and $\pi^{-}$ for the hard and supersoft symmetry energies.}
\end{figure*}

\section{IV. Conclusions}

The influence of the in-medium effects in nuclear reactions is investigated within the LQMD model, i.e., the NN cross sections, propagation of pions in nuclear medium. The in-medium cross sections are of importance on isospin emissions and lead to a flat distribution on the transverse flows and elliptic flows of free nucleons comparing with the ones in free space. The rapidity distribution of the transverse flow difference between neutron and proton depends on the density dependence of nuclear symmetry energy, in particular around the projectile rapidity. The in-medium effect and collision centrality slight affect the flow difference. The elliptic flow of free nucleons is also sensitive to the symmetry energy, especially at the high momentum in the case of the mass splitting of $m_{n}^{\ast}>m_{p}^{\ast}$. The pion optical potential plays a significant role on the transverse momentum distribution of $\pi^{-}/\pi^{+}$ ratio. The transverse flow of charged pions weakly depend on the symmetry energy and also the mass splitting. An antiflow for $\pi^{+}$ emission is pronounced in both mass splittings

\section{Acknowledgements}

Fruitful discussions with Prof. M. Di Toro, Prof. H. Lenske, Dr. M. Colonna and Dr. T. Gaitanos are acknowledged. This work was supported by the National Natural Science Foundation of China under Grant Nos 10805061 and 11175218, and the Advancement Society of Young Innovation of Chinese Academy of Sciences. The author is grateful to the support of K. C. Wong Education Foundation (KCWEF) and DAAD during his research stay in Justus-Liebig-Universit\"{a}t Giessen, Germany.

\end{document}